\documentclass[a4paper,11pt]{article}
\usepackage{jheppub} 
\usepackage[utf8]{inputenc}
\usepackage[T1]{fontenc} 
\usepackage{amsmath,amsthm,amssymb}
\usepackage{graphicx}
\usepackage{subfigure}
\usepackage{xcolor}
\usepackage{float}
\usepackage{array}
\usepackage{arydshln}
\graphicspath{{fig/}{image/}}
\usepackage{hyperref}
\hypersetup{colorlinks=true,linkcolor=blue,citecolor=magenta}
\usepackage{booktabs}
\usepackage{array}

\newcommand{\dd}{\mathrm{d}}

\newtheorem*{thm*}{Theorem}

\title{\boldmath  Central charge and black hole entropy for regular extremal black-bounce spacetimes}

\author[a,b]{Xu Ye,}
\author[c]{Shan-Ping Wu,}
\author[a,b]{Yu-Kun Zhang,}
\author[1,a,b]{and Shao-Wen Wei\note{Corresponding author.}}

\affiliation[a]{Lanzhou Center for Theoretical Physics, Key Laboratory of Theoretical Physics of Gansu Province, Key Laboratory of Quantum Theory and Applications of MoE, Lanzhou University, Lanzhou 730000, China}
\affiliation[b]{Institute of Theoretical Physics \& Research Center of Gravitation, School of Physical Science and Technology, Lanzhou University, Lanzhou 730000, China}
\affiliation[c]{ School of Electronics and Information Engineering, Shaoxing Institute of Technology, Shaoxing, 312000, China}

\emailAdd{yex2026@lzu.edu.cn}
\emailAdd{wushanping@zsit.edu.cn}
\emailAdd{ykzhang2021@lzu.edu.cn}
\emailAdd{weishw@lzu.edu.cn}

\abstract{
The Bekenstein-Hawking entropy, proportional to one quarter of the horizon area, is fundamental in black hole thermodynamics and can also be understood via the AdS/CFT correspondence, such as the 3D BTZ black hole and 2D CFT. In this work, we adopt the Kerr/CFT approach to analyze the central charge and black hole entropy for regular extremal black-bounce spacetimes, including the counterparts of the Kerr, Kerr-Newman, and Reissner-Nordstr\"om black holes. These spacetimes are free of curvature singularities at $r=0$. We derive the near horizon geometries of these spacetimes and find that they exhibit enhanced symmetry, namely SL$(2,\mathbb{R})\times \mathrm{U}(1)$ or SL$(2,\mathbb{R}) \times \mathrm{SO}(3)$. By imposing appropriate boundary conditions, we analyze their asymptotic symmetry groups, which contain diffeomorphisms as well as the $\mathrm{U}(1)_{\rm gauge}$ symmetry arising from the electromagnetic field. We then extract the central charge from the charge algebra and evaluate the left-moving temperature of the Frolov-Thorne vacuum. It is worth emphasizing that in the black-bounce Kerr-Newman case, the central charge from the electromagnetic contribution vanishes. Furthermore, in the black-bounce Reissner-Nordstr\"om case, we uplift the 4D geometry to a 5D configuration by incorporating a $\mathrm{U}(1)$ gauge fiber. Our results show that the microscopic entropy calculated from the Cardy formula is consistent with the Bekenstein-Hawking entropy. This agreement suggests that the Kerr/CFT approach remains valid for certain regular spacetimes without curvature singularities, thereby providing a microscopic statistical understanding of black hole entropy.}


\begin{document} 
\maketitle
\flushbottom

\section{Introduction} \label{sec:intro}

The Bekenstein-Hawking area law for black hole entropy underpins black hole thermodynamics and the black hole information \cite{Bekenstein:1973ur,Hawking:1975vcx,Bardeen:1973gs,Hawking:1982dh,Page:1993wv,Eisert:2008ur,Harlow:2014yka}. This discovery indicates that the entropy of a black hole system is encoded on a two-dimensional surface, namely the black hole event horizon. This insight led to the holographic principle, proposed by 't Hooft and Susskind \cite{tHooft:1993dmi,Susskind:1994vu}, which relates a quantum theory of gravity to a quantum field theory without gravity in fewer dimensions \cite{Bigatti:1999dp,Bousso:2002ju}.

When combined with string theory, this principle gives rise to a particularly powerful the AdS/CFT correspondence constructed between $D=4$, $\mathcal{N}=4$ super-Yang-Mills theory and type IIB string theory on $\mathrm{AdS}_5 \times \mathrm{S}^5$ spacetime \cite{Maldacena:1997re,Witten:1998qj,Gubser:1998bc,Aharony:1999ti}. In the context of string theory, one can account for the Bekenstein-Hawking entropy of certain extremal supersymmetric black holes from the microscopic statistics of Bogomol\'nyi--Prasad--Sommerfield (BPS) states, as shown in Ref. \cite{Strominger:1996sh}. Notably, this statistical explanation relies strongly on supersymmetry, which is essential for counting BPS states. In fact, the work of Brown and Henneaux \cite{Brown:1986nw} demonstrates that any consistent theory of quantum gravity on 3D AdS spacetime is holographically dual to a 2D CFT, a duality that need not invoke string theory or supersymmetry. Investigations of near horizon geometry and the attractor mechanism indicate that only extremality is required \cite{Strominger:1997eq,Sen:2007qy}.

Consider a non-supersymmetric but extremal black hole, the Kerr/CFT correspondence was established in Ref. \cite{Guica:2008mu}, which is a duality between 4D extremal Kerr black holes and a 2D CFT. The Bekenstein-Hawking entropy of the extremal Kerr black hole was reproduced as the statistical entropy of the dual CFT using the Cardy formula \cite{Cardy:1986ie}. More precisely, the authors examined the near horizon geometry of extremal Kerr black holes, which exhibits an enhanced $\mathrm{SL}(2,\mathbb{R}) \times \mathrm{U}(1)$ isometry group. At a fixed polar angle $\theta=\theta_0$, the corresponding geometry is a quotient of warped $\mathrm{AdS}_3$ \cite{Anninos:2008fx}. Following the $\mathrm{AdS}_3/\mathrm{CFT}_2$ argument of Brown and Henneaux \cite{Brown:1986nw}, the authors analyzed the asymptotic symmetry group under an appropriate boundary condition and found that the charge algebra acquires a central extension with central charge $c_L=12J$. The left-moving temperature $T_L=\frac{1}{2 \pi}$ of the Frolov-Thorne vacuum \cite{Frolov:1989jh} can be obtained by expanding the Boltzmann factor of the density matrix into the near horizon region. Consequently, the microscopic entropy from the Cardy formula is $S_{\rm mic}=\frac{\pi^2}{3}c_L T_L=2 \pi J$, consistent with the Bekenstein-Hawking entropy of the extremal Kerr black hole. See Refs.~\cite{Hartman:2008pb,Lu:2008jk,Compere:2009dp,Bredberg:2009pv,Castro:2010fd,Compere:2012jk,Wang:2010qv,Compere:2015bca,Hajian:2017mrf,Stepanenko:2022gwy,Volovik:2025yzr} for subsequent work on related generalizations.

In this paper, we use the Kerr/CFT correspondence to explore regular extremal black-bounce spacetimes, including the black-bounce counterparts of Kerr, Kerr--Newman, and Reissner--Nordstr\"om black holes \cite{Simpson:2018tsi,Mazza:2021rgq,Franzin:2021vnj}. Unlike the usual black hole spacetime, these black-bounce solutions are regular with no curvature singularity at $r=0$, that is, curvature invariants such as $R$, $R_{\mu \nu}R^{\mu \nu}$, $R^{\mu \nu \alpha \beta}R_{\mu \nu \alpha \beta}$, and $C^{\mu \nu \alpha \beta} C_{\mu \nu \alpha \beta}$ are all finite at $r=0$. We believe that any consistent quantum theory of gravity should resolve the curvature singularity of general relativity. Moreover, in realistic astrophysical environments, all compact objects or black holes should be regular. Therefore, studying these regular black-bounce spacetimes is important for uncovering the underlying features of quantum gravity and real astrophysical circumstances. For each extremal black-bounce spacetime, we first obtain its near horizon geometry and analyze the corresponding asymptotic symmetry group. After specifying the boundary condition, we examine the charge algebra via the Dirac bracket. We then compute the central charge and the temperature of the Frolov-Thorne vacuum. The resulting microscopic entropy agrees with the Bekenstein-Hawking entropy of the extremal black-bounce spacetime.

This paper is organized as follows. In section~\ref{sec:kerr}, we study the black-bounce Kerr spacetime, a regular rotating spacetime. We evaluate the near horizon geometry of the extremal case in the $(T,R,\theta,\phi_1)$ coordinates. Its asymptotic symmetry group consists of diffeomorphisms, and from the corresponding charge algebra we read off the central charge. We then analyze the left-moving temperature of the Frolov-Thorne vacuum and compute the microscopic entropy via the Cardy formula, which agrees with the Bekenstein--Hawking entropy. Section~\ref{sec:KN} is devoted to the black-bounce Kerr-Newman case, which includes an electromagnetic field. Following similar logic, we give its central charge, left-moving temperature, and microscopic entropy. Here the asymptotic symmetry group contains the diffeomorphisms and the $\mathrm{U}(1)_{\rm gauge}$ gauge transformation from the electromagnetic field. In section~\ref{sec:RN}, we discuss the black-bounce Reissner-Nordstr\"om black hole, a static, spherically symmetric spacetime. We combine its near horizon geometry with the $\mathrm{U}(1)$ electromagnetic field to form a 5D geometry. For this 5D geometry, we compute the microscopic entropy and find it consistent with the Bekenstein--Hawking entropy. Finally, section~\ref{sec:conclu} summarizes our results and discusses the physical implications. Throughout this work, we adopt geometrized units where $c=\hbar=1$, $G^{(4)}=1$, and $G^{(5)}=2 \pi$.

\section{Black-bounce Kerr spacetime}\label{sec:kerr}

We begin with the black-bounce Kerr spacetime, which is a regular counterpart of Kerr case. Its line element can be obtained from the Kerr metric via the replacement $r \to \sqrt{r^2+\ell^2}$ \cite{Mazza:2021rgq} 
\begin{equation}
    \dd s^2=-(1-\frac{2m \sqrt{r^2+\ell^2}}{\Sigma }) \dd t^2+ \frac{\Sigma }{\Delta} \dd r^2+\Sigma \dd \theta^2-\frac{4m a \sin^2 \theta \sqrt{r^2+\ell^2}}{\Sigma} \dd t \dd \phi +\frac{A \sin ^2 \theta }{\Sigma } \dd \phi^2,
\end{equation}
with
\begin{align}
    &\Sigma =r^2+\ell^2+a^2 \cos ^2 \theta,\\
    &\Delta=r^2+\ell^2+a^2-2m \sqrt{r^2+\ell^2}, \\
    & A=(r^2+\ell^2+a^2)^2 -\Delta a^2 \sin ^2 \theta.
\end{align}
Here $m$, $a$, $\ell$ are black hole mass, spin, and regularization parameter, respectively.

This line element can reduce to the Kerr metric when $\ell =0$ and to the Simpson-Visser metric \cite{Simpson:2018tsi} when $a=0$. One can obtain the event horizon via solving $\Delta(r_\pm) =0$
\begin{align}
    r_+=\sqrt{(m+\sqrt{m^2-a^2})^2-\ell^2}, \qquad r_-=\sqrt{(m-\sqrt{m^2-a^2})^2-\ell^2}.
\end{align}

Since the Killing vector $\chi =\partial_t+\Omega_H \partial_\phi$ of event horizon  is null, one has
\begin{align}
    g_{\mu \nu} \chi^\mu \chi ^\nu=g_{tt}+2 g_{t \phi} \Omega_H+ g_{\phi \phi} \Omega_H^2=0.
\end{align}

From this equation, we can give the angular velocity at the horizon
\begin{align}
    \Omega_H=-\frac{g_{t \phi}}{g_{\phi \phi}} \bigg|_{r=r_+}=\frac{a}{r_+^2+\ell^2+a^2}.
\end{align}

The Bekenstein-Hawking temperature and black hole entropy for this black hole are 
\begin{align}
    T_H&=\frac{\kappa}{2 \pi}=\frac{1}{2 \pi}\times \frac{\Delta' (r_+)}{2(r_+^2+a^2+\ell^2)},\\[2mm]
    S_{\rm BBK}&= \frac{1}{4} A_{H}=\frac{1}{4} \iint d \theta d \phi \sqrt{\det g} \bigg|_{r_+}= \pi(r_+^2+\ell^2+a^2) .
\end{align}
Here we used the $\sqrt{\det g } \big|_{r_+}= \sqrt{g_{\theta \theta} g_{\phi \phi}}\big|_{r_+}= \sin \theta(r_+^2+\ell^2+a^2)$.

The extremal black holes occur when $r_+=r_-$, that is the inner and outer horizons coincide. This condition is equivalent to
\begin{align} 
    a=m.
\end{align}
The radius of horizon becomes
\begin{align}
    r_e=\lim_{a \to m} r_+=\lim_{a \to m} r_-=\sqrt{m^2-\ell^2}.
\end{align}

Taking the limit $a\to m$, the line element of extremal black-bounce Kerr spacetime turns to 
\begin{equation}
    \dd s^2=-(1-\frac{2m \sqrt{r^2+\ell^2}}{\Sigma }) \dd t^2+ \frac{\Sigma }{\Delta} \dd r^2+\Sigma \dd  \theta^2-\frac{4m^2 \sin^2 \theta \sqrt{r^2+\ell^2}}{\Sigma} \dd t \dd  \phi +\frac{A \sin ^2 \theta }{\Sigma } \dd  \phi^2,
\end{equation}
with
\begin{align}
    &\Sigma =r^2+\ell^2+m^2 \cos ^2 \theta,\\
    &\Delta= (\sqrt{r^2+\ell^2}-m)^2, \\
    & A=(r^2+\ell^2+m^2)^2 -\Delta m^2 \sin ^2 \theta.
\end{align}

The angular velocity $\Omega_H$ and temperature $T_H$ for extremal black holes becomes to 
\begin{align}
    \Omega_e=\lim_{a \to m}\Omega_H=\frac{1}{2m}, \qquad T_e=\lim_{a \to m}T_H=0. 
\end{align}

The Bekenstein-Hawking entropy in the extremal situation is
\begin{align}
    S_{\rm EBBK}= \lim_{a \to m} S_{\rm BBK}= 2 m^2 \pi.
\end{align}
Here the subscript EBBK denotes the extremal black-bounce Kerr.

\subsection{Near horizon geometry}

We now consider the near horizon geometry for the extremal black-bounce Kerr spacetime. To realize this, we can use the following coordinates transformation from $(t,r,\theta,\phi)$ to $(T,R,\theta,\phi_1)$ \cite{Bardeen:1999px,Kunduri:2007vf,Kunduri:2013gce}
\begin{align}\label{eq:coord1}
    r&=r_e+ \lambda  r_0 R, \nonumber\\
    t&=r_0 r_1 \frac{T}{\lambda },\\
    \phi&=\phi_1+\frac{T r_0 r_1}{\lambda }\Omega _e. \nonumber 
\end{align}
Here $r_0=\sqrt{2}m$ and $r_1=m^2/ (m^2-\ell^2)$.

Upon the above coordinate transformation \eqref{eq:coord1}  and taking limit $\lambda \to 0$, we can obtain the corresponding near horizon geometry for black-bounce Kerr spacetime
\begin{align}\label{eq:NHGK} 
    \dd s^2_{\rm NHG}=\Gamma(\theta) \left(-  R^2\,\dd T^2 +\frac{\dd R^2}{R^2 }+\alpha^2 \dd  \theta^2 \right)+\gamma(\theta) (\dd  \phi_1+ k R \dd T)^2,
\end{align}
with
\begin{align}
    &\Gamma(\theta) = \frac{m^4 \left(\cos ^2(\theta )+1\right)}{m^2-\ell^2}, \\
    &\gamma(\theta)= \frac{8 m^2 \sin ^2(\theta )}{\cos (2 \theta )+3}, \\
    & \alpha=\frac{\sqrt{m^2-\ell^2}}{m}, \qquad  k=\frac{1}{\alpha}.
\end{align}

This metric has the enhanced SL$(2,\mathbb{R})\times \mathrm{U}(1)$ symmetry \cite{Hartman:2008pb} with generators 
\begin{align}
    &L_0=\partial_{\phi_1}. \\
    &K_{-1}=\partial_T,\quad K_{0}=T \partial_T-R \partial_R, \quad K_{+1}=(\frac{1}{2R^2}+\frac{T^2}{2})\partial_T-T R \partial_R-\frac{k}{R}\partial_{\phi_1}.
\end{align}
Note that the $\mathrm{U}(1)$ symmetry is generated by $L_0$, and the SL$(2,\mathbb{R})$ symmetry is generated by $K_{-1}$, $K_0$ and $K_{+1}$.

\subsection{Asymptotic symmetry group}

The asymptotic symmetry group (ASG) of a spacetime is the group of allowed symmetries modulo trivial symmetries \cite{Ciambelli:2022vot,Ruzziconi:2019pzd,Speziale:2025lkm}, that is
\begin{align}
    ASG=\frac{\text{Allowed symmetry}}{\text{Trivial symmetry}}.
\end{align}
A symmetry is allowed if the corresponding symmetry transformation obeys the boundary condition we specified. The trivial symmetry is refer to the symmetry transformation which leads to the vanishing conserved charge according to the Noether's theorem.

For the present case, we can choose the boundary conditions as follows \cite{Guica:2008mu}
\begin{equation}
h_{\mu \nu}=\left(
\begin{matrix}
h_{TT} = \mathcal{O}(R^2) & h_{TR} = \mathcal{O}\left(\frac{1}{R^2}\right) & h_{T\theta} = \mathcal{O}\left(\frac{1}{R}\right) & h_{T\phi_1} = \mathcal{O}(1) \\[6pt]
h_{RT} = h_{TR} & h_{RR} = \mathcal{O}\left(\frac{1}{R^3}\right) & h_{R\theta} = \mathcal{O}(\frac{1}{R^2}) & h_{R\phi_1} = \mathcal{O}(\frac{1}{R})  \\[6pt]
h_{\theta T} = h_{T\theta} & h_{\theta R} = h_{R \theta}  & h_{\theta\theta} = \mathcal{O}\left(\frac{1}{R}\right) & h_{\theta\phi_1} = \mathcal{O}(\frac{1}{R})  \\[6pt]
h_{\phi_1 T} = h_{T\phi_1} & h_{\phi_1 R} = h_{R \phi _1}  & h_{\phi_1 \theta} = h_{\theta \phi_1}  & h_{\phi_1 \phi_1} = \mathcal{O}(1) 
\end{matrix}
\right),
\end{equation}
where $h_{\mu \nu}=\delta g_{\mu \nu}$ is the metric fluctuation around the near horizon geometry \eqref{eq:NHGK}.

For the metric described by \eqref{eq:NHGK}, the symmetry transformation is the diffeomorphism. Under the above boundary condition, the most general diffeomorphisms take the form
\begin{align} \label{eq:diffeomK}
    \xi=\xi^\mu \partial_\mu=-R \epsilon'(\phi_1) \partial_R+ \epsilon (\phi_1) \partial_{\phi_1}.
\end{align}
Here we ignore the subleading terms which give the trivial symmetry transformation and not affect our discussion. The asymptotic symmetry generated by the vector field $\xi$ can be viewed as the conformal group of the circle $S^1$.

Since the coordinate $\phi _1$ has period $\phi _1 \sim  \phi _1+ 2 \pi $, we can consider the Fourier expansion for function $\epsilon (\phi_1)$ and define $\epsilon _n(\phi_1)=- e^{-in \phi_1}$, then
\begin{equation}
    \xi_n= -i n R e^{-i n \phi_1 } \partial_R  -e^{-i n \phi_1 }\partial_{\phi_1}.
\end{equation}

Using the Lie bracket of vector fields, the symmetry generators satisfy the Virasoro algebra
\begin{align}\label{eq:Lie1}
    [\xi_p,\xi_n]=-i(p-n) \xi_{p+n}.
\end{align}
Note that for $n=0$, vector field $\xi_0=-\partial_{\phi_1}$ generates the $\mathrm{U}(1)$ rotational isometry.

\subsection{Central charge and temperature}

Sofar we have obtained the diffeomorphisms \eqref{eq:diffeomK}, the next step is to construct the conserved charge $Q_{\xi}[g]$ associated with these diffeomorphism. For this purpose, we can use the covariant formalism developed by Barnich, Brandt, and Compere, based on the Refs. \cite{Barnich:2001jy,Barnich:2007bf,Barnich:2003xg,Compere:2018aar}. The infinitesimal charge differences $\delta Q_\xi[g]$ between the neighboring geometries $g_{\mu \nu}$ and $g_{\mu \nu}+h_{\mu \nu}$ are given by \cite{Compere:2018aar}
\begin{align}\label{K:kgrav}
    \delta Q_\xi[g]=\frac{1}{8 \pi G} \int_{\partial \Sigma} k^{\text{grav}}_\xi[h,g],
\end{align}
where $\partial \Sigma$ is the boundary of the spatial slice and the explicit form of $k^{\rm grav}_\xi[g,h]$ is
\begin{equation}
\begin{split}
k_{\xi}^{\text{grav}}[h,g] &= \frac{1}{4} \epsilon_{\alpha\beta\mu\nu} \bigg[ \xi^{\nu} D^{\mu} h - \xi^{\nu} D_{\sigma} h^{\mu\sigma} + \xi_{\sigma} D^{\nu} h^{\mu\sigma} \\
&\quad + \frac{1}{2} h D^{\nu} \xi^{\mu} - h^{\nu\sigma} D_{\sigma} \xi^{\mu} + \frac{1}{2} h^{\sigma\nu} (D^{\mu} \xi_{\sigma} + D_{\sigma} \xi^{\mu}) \bigg] dx^{\alpha} \wedge dx^{\beta}.
\end{split}
\end{equation}
Here $h=h^\mu_{\;\mu}$ is the trace of fluctuation metric $h_{\mu \nu}$.

Note that the charged $Q_{\partial T}$ corresponding to the symmetry transformation generated by the time translation $\partial_T$ is trivial, that is 
\begin{align}
    Q_{\partial T}[g]=0.
\end{align}
This can be viewed as a constraint, so we only focus on the nonzero charge $Q_{\xi}$. From the expression \eqref{eq:diffeomK} of $\xi$, we know it depends on the function $\epsilon(\phi_1)$.

The algebra of the ASG is the Dirac bracket algebra of the charges themselves
\begin{align}\label{eq:ChaAlg}
    \{ Q_{\xi_p}, Q_{\xi_n} \}_{\text{DB}} &=\delta_{\xi_n}Q_{\xi_p} \nonumber\\
    &= Q_{[\xi_p, \xi_n]} + \frac{1}{8\pi G} \int_{\partial\Sigma} k_{\xi_p} [\mathcal{L}_{\xi_n} g, g].
\end{align}
In the second line, the surface integral term is a central extension.

Using the formula \eqref{eq:Lie1}, the above charge algebra becomes
\begin{align}
    \{ Q_p, Q_n \}_{\text{DB}} = -i (p - n)Q_{p+n} -i \frac{c}{12} \left( p^3 - Bp \right) \delta_{p+n,0},
\end{align}
where $B$ is a constant that can be absorbed by a shift in $Q_0$. Therefore the cubic term $\frac{c}{12}p^3$ is vital for us.

Let us compute the surface integral appeared in charge algebra \eqref{eq:ChaAlg}, the result is
\begin{align}
    \frac{1}{8\pi G} \int_{\partial\Sigma} k_{\xi_p} [\mathcal{L}_{\xi_n} g, g]= \frac{-i \, m^3 p^3}{\sqrt{m^2-\ell^2 }}  -2i \, m \sqrt{ m^2 - \ell^2 } p.
\end{align}

Comparing this result with the $-i \frac{c}{12} p^3 \delta_{p+n,0}$, we can read off the central charge
\begin{align}\label{eq:CLkerr}
    c_L=\frac{12 m^3}{\sqrt{m^2-\ell^2}}.
\end{align}

Let us continue to consider the temperature of Frolov-Thorne vacuum \cite{Frolov:1989jh} for extremal black-bounce Kerr spacetime, which is well behaved in the near horizon region. This vacuum is an analog of the Hartle-Hawking vacuum for Schwarzschild case. The quantum field can be expanded in the eigenmodes of the asymptotic energy $\omega$ and angular momentum $\bar{m}$
\begin{align}
    \Phi =\sum_{\omega,\bar{m},l} \phi_{\omega \bar{m} l}\, e^{-i \omega t+i \bar{m} \phi} f_l(r,\theta).
\end{align}

After tracing over the region inside the horizon, the vacuum is a diagonal density matrix with Boltzmann weighting factor
\begin{align}
    \rho= e^{ \frac{- \hbar(\omega-\Omega_H \bar{m})}{T_H}}.
\end{align}

By means of the coordinate transformation \eqref{eq:coord1}, we can express the phase factor $e^{-i \omega t+i \bar{m} \phi}$ in quantum field $\Phi$ into the near horizon region
\begin{align}
    e^{-i \omega t+i \bar{m} \phi}=e^{-i n_R T+i n_L \phi_1},
\end{align}
with
\begin{equation}
    n_L= \bar{m},  \qquad n_R= \frac{m^2(\bar{m}-2m \omega)}{\sqrt{2}(\ell^2-m^2)\lambda}.
\end{equation}

Using $n_L$ and $n_R$, the Boltzmann factor of vacuum density matrix recasts to
\begin{align}
    e^{\frac{-\hbar(\omega-\Omega_H m)}{T_H}}= e^{-\frac{n_L}{T_L}-\frac{n_R}{T_R}} ,
\end{align}
where
\begin{align}\label{eq:TLR}
    T_L=\frac{ X Y }{4 \pi  (\sqrt{m^2-a^2}-a+m)}, \qquad T_R=\frac{m^2  X Y}{4 \sqrt{2} \pi    (\sqrt{m^2-a^2}+m) (m^2-\ell^2) \lambda }.
\end{align}
Here we denote
\begin{align*}
    X&=\sqrt{\frac{\ell^2}{a^2-2 m (\sqrt{m^2-a^2}+m)}+1},\\[2mm]
    Y&=\sqrt{2 m (m+\sqrt{m^2-a^2})-a^2}-\sqrt{2 m (m-\sqrt{m^2-a^2})-a^2}.
\end{align*}

Taking the extremal limit $a \to m$ for \eqref{eq:TLR}, we have
\begin{equation}\label{eq:TLRkerr}
    T_L= \frac{\sqrt{m^2-\ell^2}}{2m \pi}, \qquad T_R=0.
\end{equation}

\subsection{Black hole entropy}

Since the quantum states on the near horizon region are identified with those of the left-moving part of the CFT via the holographic duality. The CFT dual of the Frolov-Thorne vacuum also has same temperature as \eqref{eq:TLRkerr}. The central charge of CFT is same as the result \eqref{eq:CLkerr} appeared in the charge algebra of asymptotic symmetry group.

According to the Cardy formula \cite{Cardy:1986ie} and using \eqref{eq:TLRkerr} and \eqref{eq:CLkerr}, the microscopic entropy for a unitary CFT is
\begin{align}
    S_{\rm mic}&=\frac{\pi^2}{3}c_L T_L= \frac{\pi^2}{3}\times \frac{12m^3}{ \sqrt{m^2-\ell^2}}\times  \frac{\sqrt{m^2-\ell^2}}{2m \pi}=2m^2 \pi=S_{\rm EBBK}.
\end{align}
This result reproduces the Bekenstein-Hawking entropy of the extremal black-bounce Kerr black hole.

\section{Black-bounce Kerr-Newman spacetime} \label{sec:KN}

In this section, we examine the central charge and black hole entropy for extremal black-bounce Kerr-Newman spacetime, which is a regularized version of Kerr-Newman solution. Distinct from the last section, here we also need analyze the contribution from the electromagnetic part.

The line element of black-bounce Kerr-Newman spacetime is given by \cite{Franzin:2021vnj}
\begin{align}\label{eq:LEKN}
    \mathrm{d}s^2 &= -\frac{\Delta}{\rho^2} \left( a \sin^2\theta \, \mathrm{d}\phi - \mathrm{d}t \right)^2 + \frac{\sin^2\theta}{\rho^2} \left[ \left( r^2 + \ell^2 + a^2 \right) \mathrm{d}\phi - a \, \mathrm{d}t \right]^2 + \frac{\rho^2}{\Delta} \mathrm{d}r^2 + \rho^2 \mathrm{d}\theta^2, 
    \end{align}
where
\begin{align}
    \rho^2 &= r^2 + \ell^2 + a^2 \cos^2\theta, \qquad \Delta = r^2 + \ell^2 + a^2 - 2m \sqrt{r^2 + \ell^2 } + Q^2. 
\end{align}
The parameters $m$, $a$, $Q$, and $\ell$ are black hole mass, spin, electric charge, and regularization parameter, respectively.

The corresponding electromagnetic potential takes the form
\begin{align}
    A_\mu= \frac{-Q \sqrt{r^2+\ell^2}}{\rho^2}(1,0,0,-a \sin ^2 \theta).
\end{align}

The line element \eqref{eq:LEKN} can reduce to the Kerr-Newman solution when $\ell =0$ and to the black-bounce Kerr case when $Q=0$. One can obtain the event horizon via solving $\Delta(r_\pm) =0$
\begin{align}
    r_+&=\sqrt{2 m^2-Q^2-a^2-\ell^2+2 \sqrt{-m^2 \left(a^2-m^2+Q^2\right)}},\\
    r_-&= \sqrt{2 m^2-Q^2-a^2-\ell^2-2 \sqrt{-m^2 \left(a^2-m^2+Q^2\right)}}.
\end{align}

From the null condition $g_{tt}+2 g_{t \phi} \Omega_H+ g_{\phi \phi} \Omega_H^2=0$, one can calculate the angular velocity at the horizon
\begin{align}
    \Omega_H=-\frac{g_{t \phi}}{g_{\phi \phi}} \bigg|_{r=r_+}=\frac{a}{r_+^2+\ell^2+a^2}.
\end{align}

The Bekenstein-Hawking temperature for this spacetime is
\begin{align}
    T_H=\frac{\kappa}{2 \pi}=\frac{1}{2 \pi}\times \frac{\Delta' (r_+)}{2(r_+^2+a^2+\ell^2)}.
\end{align}

The corresponding black hole entropy is
\begin{align}
    S_{\rm BBKN}= \frac{1}{4} A_{H}=\frac{1}{4} \iint d \theta d \phi \sqrt{\det g} \bigg|_{r_+}= \pi(r_+^2+\ell^2+a^2) .
\end{align}
Here we used the $\sqrt{\det g } \big|_{r_+}= \sqrt{g_{\theta \theta} g_{\phi \phi}}\big|_{r_+}= \sin \theta(r_+^2+\ell^2+a^2)$.

When $r_+=r_-$, one can get the extremal black-bounce Kerr-Newman black holes. Solving $r_+=r_-$ yields the relation
\begin{align} 
    a^2=m^2-Q^2.
\end{align}

Taking the extremal limit $a\to \sqrt{m^2-Q^2}$, the horizon becomes
\begin{align}
    r_e=\lim_{a \to \sqrt{m^2-Q^2}} r_+= \lim_{a \to \sqrt{m^2-Q^2}} r_-=\sqrt{m^2-\ell^2}. 
\end{align}

In this extremal limit, the black-bounce Kerr-Newman line element recasts to 
\begin{align}
    \mathrm{d}s^2 =& -\frac{\Delta}{\rho^2} \left(\sqrt{m^2-Q^2}  \sin^2\theta \, \mathrm{d}\phi - \mathrm{d}t \right)^2 \nonumber\\
    & + \frac{\sin^2\theta}{\rho^2} \left[ \left( r^2 + \ell^2 + m^2-Q^2 \right) \mathrm{d}\phi - \sqrt{m^2-Q^2}\,  \mathrm{d}t \right]^2 + \frac{\rho^2}{\Delta} \mathrm{d}r^2 + \rho^2 \mathrm{d}\theta^2, 
\end{align}
where
\begin{align}
    \rho^2 &= r^2 + \ell^2 + (m^2-Q^2) \cos^2\theta, \qquad \Delta =(\sqrt{r^2+\ell^2}-m)^2. 
\end{align}

At the same tme, the electromagnetic potential $A_\mu$ in this limit is written as 
\begin{align}
    A_\mu= \frac{-Q \sqrt{r^2+\ell^2}}{\rho^2}(1,0,0,-\sqrt{m^2-Q^2} \sin ^2 \theta).
\end{align}

The angular velocity and Bekenstein-Hawking temperature at the horizon for extremal situation turn to
\begin{align}
    \Omega_e=\lim_{a \to \sqrt{m^2-Q^2}} \Omega_H&=\frac{\sqrt{m^2-Q^2}}{2m^2-Q^2},\\
    T_e=\lim_{a \to \sqrt{Q^2-m^2}} T_H&=0.
\end{align}

The entropy of extremal black-bounce Kerr-Newman black holes is
\begin{align}
    S_{\rm EBBKN}=\lim_{a \to \sqrt{m^2-Q^2}} S_{\rm BBK}=  \pi(2 m^2-Q^2).
\end{align}

\subsection{Near horizon geometry}

Now we consider the near horizon geometry for the extremal black-bounce Kerr-Newman spacetime. To do this, we use the coordinate transformation from $(t,r,\theta,\phi)$ to $(T,R,\theta,\phi_1)$ \cite{Bardeen:1999px}
\begin{align}\label{KN:CT}
    r&=r_e+ \lambda r_0 R, \nonumber\\
    t&=r_0 r_1 \frac{T}{\lambda },\\
    \phi&=\phi_1+\frac{T r_0 r_1}{\lambda }\Omega _e \nonumber.
\end{align}
Here $r_0=\sqrt{2}m$ and $r_1=m^2/ (m^2-\ell^2)$.

Upon the coordinate transformation \eqref{KN:CT} and taking the limit $\lambda \to 0$, we can obtain the corresponding near horizon geometry 
\begin{align} \label{KN:NHG}
   \mathrm{d} s^2_{\rm NHG}=\Gamma(\theta) \left(-  R^2\,\dd T^2 +\frac{\dd R^2}{R^2 }+\alpha^2 \dd \theta^2 \right)+\gamma(\theta) (\dd \phi_1+ k R \dd T)^2,
\end{align}
with
\begin{align}
    \Gamma(\theta) &=\frac{m^2 \left(\cos ^2\theta \left(m^2-Q^2\right)+m^2\right)}{m^2-\ell^2},\\ 
    \gamma(\theta)&=\frac{\sin ^2\theta \left(Q^2-2 m^2\right)^2}{\cos ^2\theta \left(m^2-Q^2\right)+m^2}, \\
    \alpha&=\frac{\sqrt{m^2-\ell^2}}{m},\\
    k&= \frac{2 m^2 \sqrt{m^2-Q^2}}{\sqrt{m^2-\ell^2} \left(2 m^2-Q^2\right)}.
\end{align}

This metric has the enhanced SL$(2,\mathbb{R})\times \mathrm{U}(1)$ symmetry. The $\mathrm{U}(1)$ symmetry is generated by 
\begin{align}
    L_0=\partial_{\phi_1}.
\end{align}
The SL$(2, \mathbb{R})$ symmetry is generated by $K_{-1}$, $K_0$, and $K_{+1}$ as follows
\begin{align}
    K_{-1}=\partial_T,\quad K_{0}=T \partial_T-R \partial_R, \quad K_{+1}=(\frac{1}{2R^2}+\frac{T^2}{2})\partial_T-T R \partial_R-\frac{k}{R}\partial_{\phi_1}.
\end{align}

For the electromagnetic potential $A_\mu$, we also need perform the the coordinate transformation \eqref{KN:CT}. The resulting potential in the near horizon region takes the form 
\begin{align}\label{KN:A}
    A=A_\mu \dd x^\mu&= f(\theta)(\dd \phi_1+k R \dd T)+\beta \dd \phi_1,
\end{align}
with
\begin{align}
    \beta&=-\frac{Q^3}{2 m \sqrt{m^2-Q^2}},\\  
    f(\theta)&= -\frac{Q \left(Q^2-2 m^2\right) \left(\cos (2 \theta ) \left(Q^2-m^2\right)+m^2+Q^2\right)}{2 m \sqrt{m^2-Q^2} \left(\cos (2 \theta ) \left(m^2-Q^2\right)+3 m^2-Q^2\right)}.
\end{align}
Note that to obtain the potential \eqref{KN:A} we have used the gauge transformation $A'_\mu= A_\mu+d \Lambda$ to remove the singularity of the coordinate transformation in the limit $\lambda \to 0$. The $\Lambda $ we used is
\begin{align}
    \Lambda =\frac{ m^3 Q}{\sqrt{2 m^2-Q^2} \left(\ell^2  -m^2  \right)} \frac{T}{\lambda }. 
\end{align}

\subsection{Asymptotic symmetry group}

For the present case, the boundary conditions for the metric fluctuation $h_{\mu \nu}$ around the near horizon geometry are chosen to be 
\begin{equation} \label{KN:bcG}
    h_{\mu \nu}=\left(
\begin{matrix}
h_{TT} = \mathcal{O}(R^2) & h_{TR} = \mathcal{O}\left(\frac{1}{R^2}\right) & h_{T\theta} = \mathcal{O}\left(\frac{1}{R}\right) & h_{T\phi_1} = \mathcal{O}(1) \\[6pt]
h_{RT} = h_{TR} & h_{RR} = \mathcal{O}\left(\frac{1}{R^3}\right) & h_{R\theta} = \mathcal{O}(\frac{1}{R^2}) & h_{R\phi_1} = \mathcal{O}(\frac{1}{R})  \\[6pt]
h_{\theta T} = h_{T\theta} & h_{\theta R} = h_{R \theta}  & h_{\theta\theta} = \mathcal{O}\left(\frac{1}{R}\right) & h_{\theta\phi_1} = \mathcal{O}(\frac{1}{R})  \\[6pt]
h_{\phi_1 T} = h_{T\phi_1} & h_{\phi_1 R} = h_{R \phi _1}  & h_{\phi_1 \theta} = h_{\theta \phi_1}  & h_{\phi_1 \phi_1} = \mathcal{O}(1)
\end{matrix}
\right).
\end{equation}

The symmetry transformation for the near horizon metric \eqref{KN:NHG} is the diffeomorphism. Under this diffeomorphism, one has
\begin{equation}
    \delta_\xi g_{\mu \nu}=\mathcal{L}_\xi g_{\mu \nu}.
\end{equation}
If $\xi$ is an exact Killing vector, then $\delta_\xi g_{\mu \nu}=0$ gives the Killing equation, which implies the symmetry generated by $\xi$ is an exact symmetry.

Under the boundary condition \eqref{KN:bcG}, the most general diffeomorphism takes the form
\begin{align}
    \xi=\xi^\mu \partial_\mu=-R \epsilon'(\phi_1) \partial_R+ \epsilon (\phi_1) \partial_{\phi_1}.
\end{align}
The asymptotic symmetry generated by the vector field $\xi$ is a conformal group of the circle $S^1$.

Since $\phi _1 \sim  \phi _1+ 2 \pi$, we can consider the Fourier expansion for function $\epsilon (\phi_1)$ and define $\epsilon _n(\phi_1)=- e^{-in \phi_1}$, then
\begin{equation}
    \xi_n= -i n R e^{-i n \phi_1 } \partial_R  -e^{-i n \phi_1 }\partial_{\phi_1}.
\end{equation}

Using the Lie bracket of vector fields, the symmetry generators satisfy the Virasoro algebra
\begin{align}
    [\xi_p,\xi_n]=-i(p-n) \xi_{p+n}.
\end{align}

Let us consider the symmetry transformation for the electromagnetic potential $A_\mu$. The asymptotic symmetry from the Maxwell theory includes the diffeomorphism and $\mathrm{U}(1)$ gauge transformation such as
\begin{align}
    &\delta_\xi A_\mu=\mathcal{L}_\xi A_\mu,\\
    & \delta_\Lambda A=d \Lambda. 
\end{align}

For the gauge field $A_\mu$, we also need impose the boundary condition \cite{Hartman:2008pb}
\begin{align}\label{KN:bcA}
    a_\mu \sim \left(\mathcal{O}(R), \mathcal{O}(\frac{1}{R^2}), \mathcal{O}(1), \mathcal{O}(\frac{1}{R})\right).
\end{align}
Here $a_\mu=\delta A_\mu$ is the fluctuation of the background field $A_\mu$ \eqref{KN:A}, which is same as that $h_{\mu \nu}=\delta g_{\mu \nu}$ is the metric fluctuation for near horizon geometry \eqref{KN:NHG}.

Combining the diffeomorphism and $\mathrm{U}(1)$ gauge transformation, one has
\begin{align}
    \delta A&= \delta_\xi A+\delta_\Lambda  A = \mathcal{L}_\xi A_\mu+\dd \Lambda \\
        &=-kR f(\theta) \epsilon'(\phi_1) \dd T-f'(\theta) \epsilon (\phi_1) \dd \theta. \nonumber
\end{align}
Here we take $\Lambda =-f(\theta)\epsilon(\phi_1)$. Upon Fourier expansion, $\Lambda _n=f(\theta) e^{-in \phi_1}$ with respect to the basis $\epsilon_n =- e^{-in \phi_1}$.

With above boundary conditions \eqref{KN:bcG} and \eqref{KN:bcA} for $h_{\mu \nu}$ and $a_\mu$, the asymptotic symmetries consist of the pair $(\xi_n,\Lambda_n)$. The corresponding algebra is
\begin{align}\label{eq:Lie2}
    [(\xi_n,\Lambda _n),(\xi_p,\Lambda _p)]=([\xi_n,\xi_p], [\Lambda _n,\Lambda _p]_\xi),
\end{align}
where $[\xi_n,\xi_p]$ is the Lie bracket same as \eqref{eq:Lie1}. The commutator $[\Lambda_n,\Lambda _p]_\xi$ is defined as
\begin{align}
    [\Lambda _n,\Lambda _p]_\xi:= \xi_n^\mu \partial_\mu \Lambda _p-\xi^\mu_p \partial_\mu \Lambda _n.
\end{align}

As a result, the asymptotic symmetry algebra \eqref{eq:Lie2} turns to
\begin{align}\label{KN:LB2}
    [(\xi_n, \Lambda_n), (\xi_p, \Lambda_p)] =-i(n - p)(\xi_{n+p}, \Lambda_{n+p}).
\end{align}
This is the Virasoro algebra with vanishing central extension.

\subsection{Central charge and temperature }

After analyzing the diffeomorphism and guage transformation for gravity and electromagnetic part, we will construct the conserved charge $Q_{\xi,\Lambda }$ associated with asymptotic symmetry transformation. For the black-bounce Kerr-Newman spacetime, the infinitesimal charge $\delta Q_{\xi,\Lambda}$ is given by \cite{Compere:2009dp}
\begin{align}
    \delta Q_{\xi,\Lambda}=\frac{1}{8 \pi G} \int \left( k^{\text{grav}}_\xi[h;g]+k_{\xi,\Lambda }^{\text{gauge}}[h,a;g,A]\right).
\end{align}
where $k_\xi^{\text{grav}}$ can be found in \eqref{K:kgrav}, and $k_{\xi,\Lambda }^{\text{gauge}}$ is expressed as
\begin{equation}
\begin{aligned}
    k_{\xi, \Lambda}^{\text{gauge}}[h,a;g,A] =& \frac{1}{8} \epsilon_{\alpha\beta\mu\nu} \bigg[ \left( -\frac{1}{2}hF^{\mu\nu} + 2F^{\mu\gamma}h_{\gamma}^{\,\,\nu} - \delta F^{\mu\nu} \right) (\xi^\rho A_\rho + \Lambda)\\
    & - F^{\mu\nu}\xi^\rho a_\rho - 2F^{\alpha\mu}\xi^\nu a_\alpha-a^\mu g^{\nu \rho}(\mathcal{L}_\xi A_\rho + \partial_\rho \Lambda) \bigg] dx^\alpha \wedge dx^\beta.
\end{aligned}
\end{equation}
Here $\delta F^{\mu\nu} \equiv g^{\mu\alpha}g^{\nu\beta}(\partial_\alpha a_\beta - \partial_\beta a_\alpha)$. 

Note that the charges $Q_{\partial T}$ and $Q_{\Lambda }$ corresponding to time translation $\partial_T$ and gauge transformation $\Lambda(T,\theta)$ are trivial, that is 
\begin{align}
    Q_{\partial T}[g]=0, \qquad Q_{\Lambda}=0.
\end{align}
The reason is that $\partial_T$ and $\Lambda(T,\theta)$ commute with other generators in the ASG.

The algebra of the ASG is the Dirac bracket algebra between $Q_{\xi,\Lambda }$ and $Q_{\tilde{\xi}, \tilde{\Lambda }}$ 
\begin{align}\label{KN:CA}
    \{ Q_{\xi,\Lambda }, Q_{\tilde{\xi}, \tilde{\Lambda }} \}_{\text{DB}} &= (\delta_{\tilde{\xi}}+\delta_{\tilde{\Lambda }}) Q_{\xi,\Lambda } \\
    &= Q_{[(\xi,\Lambda ),(\tilde{\xi},\tilde{\Lambda })]} + \frac{1}{8\pi G} \int \left( k^{\text{grav}}_{\xi} [\mathcal{L}_{\tilde{\xi}} g, g]+k_{\xi,\Lambda }^{\text{gauge}}[\mathcal{L}_{\tilde{\xi}}g,\mathcal{L}_{\tilde{\xi}} A+d \tilde{\Lambda };g,A] \right) \nonumber
\end{align}
Here $g$ and $A_\mu$ are the near horizon geometry metric \eqref{KN:NHG} and electromagnetic potential \eqref{KN:A}. Moreover, the surface integral term is a central extension.

Using the formula \eqref{KN:LB2}, the above charge algebra becomes
\begin{align}\label{KN:DB}
    \{ Q_p, Q_n \}_{\text{DB}} = -i (p - n)Q_{p+n} -i \frac{c}{12} \left( p^3 - Bp \right) \delta_{p+n,0},
\end{align}
where constant $B$ can be absorbed by a shift in $Q_0$, and we can extract the central charge from the cubic term $\frac{c}{12}p^3$. 

Let us compute the surface integral of gravity part appeared in charge algebra \eqref{KN:CA}, the result is
\begin{align}
    \frac{1}{8\pi G} \int k^{\rm  grav}_{\xi_p} [\mathcal{L}_{\xi_n} g, g]= \frac{-i \, m^2 \sqrt{m^2-Q^2} }{\sqrt{m^2-\ell^2 }}p^3+ \mathcal{O}(p) .
\end{align}

Comparing this result with the $-i \frac{c}{12} p^3 \delta_{p+n,0}$, we can read off the central charge of gravitational contribution
\begin{align}
    c_{\rm grav}= \frac{12 m^2 \sqrt{m^2-Q^2}}{\sqrt{m^2-\ell^2}}.
\end{align}

On the other hand, the central charge from the electromagnetic contribution can be computed as
\begin{align}
    \frac{1}{8\pi G} \int  k_{\xi_p,\Lambda_p }^{\text{gauge}}[\mathcal{L}_{\tilde{\xi}_n}g,\mathcal{L}_{\tilde{\xi}_n} A+d \tilde{\Lambda}_n;g,A]=\mathcal{O}(p),\quad \Rightarrow \quad  c_{\text{gauge}}=0.
\end{align}
The result $c_{\text{gauge}}=0$ is not too surprising, since in the Kerr-Newman spacetime the central charge of Maxwell part also vanishes \cite{Hartman:2008pb}.

Consequently, the total central charge is
\begin{align}\label{KN:CL}
    c_L=c_{\text{grav}}+c_{\text{gauge}}=\frac{12 m^2 \sqrt{m^2-Q^2}}{\sqrt{m^2-\ell^2}}.
\end{align}

Let us consider the temperature of Frolov-Thorne vacuum \cite{Frolov:1989jh} for black-bounce Kerr-Newman spacetime, which is well behaved in the near horizon region. After tracing over the region inside the horizon, the vacuum is a diagonal density matrix with Boltzmann factor
\begin{align}
    \rho= e^{ \frac{- \hbar(\omega-\Omega_H \bar{m})}{T_H}}.
\end{align}

By means of the coordinate transformation \eqref{KN:CT}, the phase factor $e^{-i \omega t+i \bar{m} \phi}$ can be expressed into the near horizon region
\begin{align}
    e^{-i \omega t+i \bar{m} \phi}=e^{-i n_R T+i n_L \phi_1},
\end{align}
with
\begin{equation}
    n_L= \bar{m},  \qquad n_R= -\frac{m^2 \left(\bar{m} \sqrt{m^2-Q^2}+\omega  \left(Q^2-2 m^2\right)\right)}{\lambda \left(m^2-\ell^2\right) \sqrt{2 m^2-Q^2}}. 
\end{equation}

Using $n_L$ and $n_R$, the Boltzmann factor of vacuum density matrix recasts to
\begin{align}
    e^{\frac{-\hbar(\omega-\Omega_H \bar{m})}{T_H}}= e^{-\frac{n_L}{T_L}-\frac{n_R}{T_R}} ,
\end{align}
where
\begin{align}
    T_L=&-\frac{\left(2 m^2-Q^2\right) X }{4 \pi  \left(\sqrt{m^2-Q^2} \left(2 Y +2 m^2-Q^2\right)+a \left(Q^2-2 m^2\right)\right)} \nonumber \\
    &\times \bigg[\sqrt{-2Y -a^2+2 m^2-Q^2}-\sqrt{2Y -a^2+2 m^2-Q^2}\bigg],\\[2mm]
    T_R=&-\frac{m^2 \sqrt{2 m^2-Q^2} X }{4 \pi \lambda    \left(m^2-L^2\right) \left(2 Y+2 m^2-Q^2\right)} \nonumber \\
    &\times \bigg[\sqrt{-2 Y-a^2+2 m^2-Q^2}-\sqrt{2Y -a^2+2 m^2-Q^2}\bigg].
\end{align}
Here we denote $X$  and $Y$ as 
\begin{align*}
    X&=\sqrt{\frac{\ell^2}{-2 \sqrt{-m^2 \left(a^2-m^2+Q^2\right)}+a^2-2 m^2+Q^2}+1}, \\[2mm]
    Y&=\sqrt{-m^2 \left(a^2-m^2+Q^2\right)}.
\end{align*}
Taking the extremal limit $a \to \sqrt{m^2-Q^2}$, we have
\begin{equation}\label{KN:TL}
    T_L=\frac{\sqrt{m^2-\ell^2} \left(2 m^2-Q^2\right)}{4 \pi  m^2 \sqrt{m^2-Q^2}}, \qquad T_R=0.
\end{equation}

\subsection{Black hole entropy}

According to the holographic duality, the quantum states on near horizon region are identified with those of the left-moving part of the CFT. The central charge and temperature of CFT dual to the extremal black-bounce Kerr-Newman black holes are given by \eqref{KN:CL} and \eqref{KN:TL}.

Using the Cardy formula \cite{Cardy:1986ie}, the microscopic entropy of CFT can be obtained 
\begin{align}
    S_{\rm  mic}&=\frac{\pi^2}{3}c_L T_L= \frac{\pi^2}{3}\times \frac{12m^2 \sqrt{m^2-Q^2}}{ \sqrt{m^2-\ell^2}}\times  \frac{\sqrt{m^2-\ell^2}(2m^2-Q^2)}{4 \pi m^2 \sqrt{m^2-Q^2}} \nonumber \\
    &= \pi(2m^2-Q^2)=S_{\rm EBBKN}.
\end{align}
This result reproduce the Bekenstein-Hawking entropy of the extremal black-bounce Kerr-Newman black hole.

\section{Black-bounce Reissner-Nordstr\"om spacetime} \label{sec:RN}

In this section, we examine the central charge and black hole entropy for extremal black-bounce Reissner-Nordstr\"om (black-bounce RN) spacetime, which is a static spherical spacetime without the rotation. The relating works on RN/CFT corresponding refer to the refs. \cite{Garousi:2009zx,Chen:2009ht,Chen:2013rb}.

The line element for this spacetime can be obtained by taking $a\to 0$ for black-bounce Kerr-Newman case \cite{Franzin:2021vnj}
\begin{align}\label{BBRN}
    \mathrm{d}s^2 &=-f(r) \dd t^2+ \frac{\dd r^2}{f(r)}+(r^2+\ell^2) \dd \theta^2+(r^2+\ell^2) \sin ^2 \theta \dd \phi^2,
\end{align}
with
\begin{align}
    f(r)=1-\frac{2m}{\sqrt{r^2+\ell^2}}+\frac{Q^2}{r^2+\ell^2}.
\end{align}

The corresponding electromagnetic potential takes the form
\begin{align}
    A_\mu= \frac{-Q }{\sqrt{r^2+\ell^2}}(1,0,0,0).
\end{align}

The line element \eqref{BBRN} can reduce to the RN metric when $\ell =0$ and to the Simpson-Visser metric \cite{Simpson:2018tsi} when $Q=0$. Via evaluating $f(r_\pm)=0$, one can obtain the locations of event horizon 
\begin{align}
    r_+&= \sqrt{2 m^2-Q^2-\ell^2+2 m \sqrt{m^2-Q^2}},\\ 
    r_-&= \sqrt{2 m^2-Q^2-\ell^2-2 m \sqrt{m^2-Q^2}}.
\end{align}

The Bekenstein-Hawking temperature for this black hole is
\begin{align}\label{RN:TH}
    T_H=\frac{\kappa}{2 \pi}=\frac{f'(r_+)}{4 \pi}= -\frac{r_+ \left(Q^2-m \sqrt{\ell^2+r_+^2}\right)}{2 \pi  \left(\ell^2+r_+^2\right)^2}.
\end{align}

The black hole entropy is
\begin{align}\label{RN:SBH}
    S_{\rm  BBRN}= \frac{1}{4} A_{H}=\frac{1}{4} \iint d \theta d \phi \sqrt{\det g} \big|_{r_+}= \pi(r_+^2+\ell^2) .
\end{align}
Here we used the $\sqrt{\det g } \big|_{r_+}= \sqrt{g_{\theta \theta} g_{\phi \phi}}\big|_{r_+}= \sin \theta(r_+^2+\ell^2)$.

The extremal black holes occur when $r_+=r_-$, which is equivalent to the following condition
\begin{align} 
    m^2=Q^2.
\end{align}

The horizon of the extremal black holes becomes
\begin{align}
    r_e=\lim_{m \to Q}  r_+=\lim_{m \to Q} r_-=\sqrt{Q^2-\ell^2}.
\end{align}

Taking the limit $m \to Q$, the extremal black-bounce RN line element can be expressed as

\begin{align}
    \mathrm{d}s^2 &=-f(r) \dd t^2+ \frac{\dd r^2}{f(r)}+(r^2+\ell^2) \dd \theta^2+(r^2+\ell^2) \sin ^2 \theta \dd \phi^2,\\
    f(r)&=(1-\frac{Q}{\sqrt{r^2+\ell^2}})^2.
\end{align}

The electromagnetic potential $A_\mu$ in the extremal situation is written as 
\begin{align}
    A_\mu= \frac{-Q }{\sqrt{r^2+\ell^2}}(1,0,0,0).
\end{align}

The Bekenstein-Hawking temperature and entropy of extremal black hole turn to
\begin{align}
    T_e&=\lim_{m \to Q}  T_H=0,\\
    S_{\rm  EBBRN}&=\lim_{m \to Q}  S_{\rm  BBRN}= \pi Q^2. \label{RN:ES}
\end{align}

\subsection{Near horizon geometry}

Let us analyze the near horizon geometry for the extremal black-bounce RN black holes. One can use the following coordinates transformation to give the near horizon metric \cite{Compere:2012jk}
\begin{equation}\label{RN:TC}
\begin{split}
    r&=r_e+ \lambda  R,\\
    t&= r_1 \frac{T}{\lambda }. 
\end{split}
\end{equation}
Here $r_1=Q^4/ (Q^2-\ell^2)$.

After the coordinate transformation \eqref{RN:TC} and taking the limit $\lambda \to 0$, the corresponding near horizon geometry is
\begin{align}\label{RN:NHG} 
   \mathrm{d}s^2_{\rm  NHG}=v_1 \left(- R^2\,\dd T^2 +\frac{\dd R^2}{R^2 }\right) + v_2 \left( \dd \theta^2+ \sin ^2 \theta \dd \phi^2 \right),
\end{align}
with
\begin{align}
    v_1=\frac{Q^4}{Q^2-\ell^2}, \qquad v_2= Q^2.
\end{align}
Recall that in the previous black-bounce Kerr and  Kerr-Newman cases, the near horizon geometry has enhanced symmetry SL$(2,\mathbb{R})\times \mathrm{U}(1)$ where $\mathrm{U}(1)$ symmetry corresponds to the axisymmetric symmetry $\partial_{\phi_1}$. Nevertheless, in the black-bounce RN case this enhanced symmetry becomes SL$(2,\mathbb{R}) \times \mathrm{SO}(3 )$. The $\mathrm{SO}(3)$ symmetry comes from the spherical symmetry of spacetime.

The generators of the SL$(2,\mathbb{R})$ are given by
\begin{align}
    &K_{-1}=\partial_T,\quad K_{0}=T \partial_T-R \partial_R, \quad K_{+1}=\left(\frac{1}{2R^2}+\frac{T^2}{2}\right)\partial_T-T R \partial_R.
\end{align}

The generators of $\mathrm{SO}(3)$ symmetry consist of 
\begin{align}
    L_x &= \sin\phi \, \partial_\theta + \cot\theta \cos\phi \, \partial_\phi, \\
    L_y &= -\cos\phi \, \partial_\theta + \cot\theta \sin\phi \, \partial_\phi, \\
    L_z &= \partial_\phi.
\end{align}

Performing the coordinate transformation, the potential $A_\mu$ in the near horizon region takes the form 
\begin{align}\label{RN:A}
    A=A_\mu \dd x^\mu&= \beta R dT, \qquad \beta= \frac{Q^2}{\sqrt{Q^2-\ell^2}}.
\end{align}
Note that to obtain the potential \eqref{RN:A} we have used the gauge transformation $A'_\mu= A_\mu+d \Lambda$ to remove the singularity of the coordinate transformation when taking $\lambda  \to 0$. The explicit form of $\Lambda $ is
\begin{align}
    \Lambda =\frac{Q^4}{Q^2-\ell^2} \frac{T}{\lambda }. 
\end{align}

It is worth emphasizing that the electromagnetic field $A_\mu$ admits additional $\mathrm{U}(1)_{\text{gauge}}$ symmetry. We can combine this $\mathrm{U}(1)$ gauge bundle \eqref{RN:A} with the above near horizon geometry \eqref{RN:NHG} such that we finally obtain the 5D total space \cite{Hartman:2008pb,Garousi:2009zx}
\begin{align}\label{RN:5D}
    \dd s^2=\dd s^2_{\rm  NHG}+(\dd y+A)^2 ,
\end{align}
where $y$ is the fiber coordinate with period $2 \pi$, and $ds^2_{\rm  NHG}$ is the 4D near horizon metric \eqref{RN:NHG}. This treatment is same as that Kaluza-Klein reduction of 5D metric results in the 4D geometry and $\mathrm{U}(1)$ electromagnetic field.

\subsection{Asymptotic symmetry group}

Since the period coordinate $y$ describes a $S^1$ geometry, it allows us to extend the $\mathrm{U}(1)_{\text{gauge}}$ symmetry to a Virasoro algebra generated by
\begin{align}
    \xi^{(y)}=\xi^{(y)\mu}\partial_\mu=- R \epsilon'(y) \partial_R+\epsilon(y) \partial_y.
\end{align}

For the symmetry generated by vector field $\xi^{(y)}$, the corresponding charges are given by the 5D generalization of \eqref{K:kgrav}. The infinitesimal charges are \cite{Compere:2009dp}
\begin{align}
    \delta Q_{\xi^{(y)}}[g]=\frac{1}{8 \pi G^{(5)}} \int k^{\text{grav}}_{\xi^{(y)}}[h,g],
\end{align}
here 5D Newton's constant $G^{(5)}=2 \pi$ and $k^{\rm grav}_{\xi^{(y)}}=k^{\rm grav}_{\xi} \big|_{\xi=\xi^{(y)}}$ is
\begin{equation}
\begin{split}
k_{\xi}^{\text{grav}}[h,g] &= \frac{1}{4} \epsilon_{\alpha\beta\mu\nu} \bigg[ \xi^{\nu} D^{\mu} h - \xi^{\nu} D_{\sigma} h^{\mu\sigma} + \xi_{\sigma} D^{\nu} h^{\mu\sigma} \\
&\quad + \frac{1}{2} h D^{\nu} \xi^{\mu} - h^{\nu\sigma} D_{\sigma} \xi^{\mu} + \frac{1}{2} h^{\sigma\nu} (D^{\mu} \xi_{\sigma} + D_{\sigma} \xi^{\mu}) \bigg] dx^{\alpha} \wedge dx^{\beta}.
\end{split}
\end{equation}
Note that here $g_{\mu \nu}$ is the 5D geometry \eqref{RN:5D} and $h_{\mu \nu}=\delta g_{\mu \nu}$ is the metric fluctuation of $g_{\mu \nu}$.  

The boundary conditions for 5D metric fluctuation $h_{\mu \nu}$ are chosen as 
\begin{equation}\label{RN:bc}
   h_{\mu \nu}= \left(
\begin{matrix}
h_{TT} = \mathcal{O}(R^2) & h_{TR} = \mathcal{O}\left(\frac{1}{R^2}\right) & h_{T\theta} = \mathcal{O}\left(\frac{1}{R}\right) & h_{T\phi} = \mathcal{O}(1) & h_{Ty}=\mathcal{O}(1) \\[6pt]
h_{RT} = h_{TR} & h_{RR} = \mathcal{O}\left(\frac{1}{R^3}\right) & h_{R\theta} = \mathcal{O}(\frac{1}{R^2}) & h_{R\phi} = \mathcal{O}(\frac{1}{R})& h_{Ry}=\mathcal{O}(\frac{1}{R})  \\[6pt]
h_{\theta T} = h_{T\theta} & h_{\theta R} = h_{R \theta}  & h_{\theta\theta} = \mathcal{O}\left(\frac{1}{R}\right) & h_{\theta\phi} = \mathcal{O}(\frac{1}{R}) & h_{\theta y}=\mathcal{O}(\frac{1}{R})  \\[6pt]
h_{\phi T} = h_{T\phi} & h_{\phi R} = h_{R \phi }  & h_{\phi \theta} = h_{\theta \phi}  & h_{\phi \phi} = \mathcal{O}(1)& h_{\phi y}=\mathcal{O}(1) \\[6pt]
h_{yT}=h_{Ty} & h_{yR}=h_{Ry} & h_{y \theta}=h_{\theta y} & h_{y \phi}=h_{\phi y} & h_{yy}=\mathcal{O}(1)
\end{matrix}
\right).
\end{equation}

The most general difeomorphisms which preserve this boundary condition are of the form
\begin{align}
    \zeta=\zeta^\mu \partial_\mu= b_T \partial_T -R \epsilon'(y) \partial_R+b_\phi \partial_\phi+ \epsilon(y) \partial_y,
\end{align}
where $b_T$ and $b_\phi$ are arbitrary constants. Furthermore, we can find
\begin{align}
    \xi^{(y)}=\zeta \big|_{b_T=0,b_\phi =0}.
\end{align}
Consequently, the vector field $ \xi^{(y)}$ preserves the above boundary condition \eqref{RN:bc}.

The symmetry transformation for 5D near horizon geometry is the diffeomorphism. Under this diffeomorphism, one has
\begin{equation}
    \delta_{\xi^{(y)}} g_{\mu \nu}=\mathcal{L}_{\xi^{(y)}} g_{\mu \nu}.
\end{equation}

Given that the coordinate $y$ has period $y \sim y+ 2 \pi$, we can consider the Fourier expansion for function $\epsilon (y)$ and define $\epsilon _n(y)=- e^{-in y}$, then
\begin{equation}
    \xi^{(y)}_n= -i n R e^{-i n y} \partial_R  -e^{-i n y }\partial_y.
\end{equation}

Using the Lie bracket of vector fields, the symmetry generators satisfy the Virasoro algebra
\begin{align}\label{RN:Lie}
    [\xi^{(y)}_p,\xi^{(y)}_n]=-i(p-n) \xi^{(y)}_{p+n}.
\end{align}
Note that for $n=0$, the $\xi^{(y)}_0=-\partial_y$ generates the $\mathrm{U}(1)_{\rm  gauge}$ symmetry of electromagnetic field $A_\mu$.

\subsection{Central charge and temperature}\label{sec:RNct}

The algebra of the ASG is the Dirac bracket algebra of the charges themselves
\begin{align}\label{RN:CA}
    \{ Q_{\xi_p^{(y)}}, Q_{\xi_n^{(y)}} \}_{\text{DB}} &=\delta_{\xi_n^{(y)}}Q_{\xi_p^{(y)}} \nonumber\\
    &= Q_{[\xi_p^{(y)}, \xi_n^{(y)}]} + \frac{1}{8\pi G^{(5)}} \int_{\partial\Sigma} k_{\xi_p^{(y)}} [\mathcal{L}_{\xi_n^{(y)}} g, g].
\end{align}
The surface integral term in the second line is a central extension.

Using the formula \eqref{RN:Lie}, the above charge algebra becomes
\begin{align}
    \{ Q_p, Q_n \}_{\text{DB}} = -i (p - n)Q_{p+n} -i \frac{c}{12} \left( p^3 - Bp \right) \delta_{p+n,0},
\end{align}
where constant $B$ can be absorbed by a shift in $Q_0$. Furthermore, we can extract the central charge from the cubic term $\frac{c}{12}p^3$. 

Let us compute the surface integral appeared in charge algebra \eqref{RN:CA}, the result is
\begin{align}
    \frac{1}{8\pi G^{(5)}} \int_{\partial\Sigma} k_{\xi_p^{(y)}} [\mathcal{L}_{\xi_n^{(y)}} g, g]= \frac{-i  Q^4}{2 \sqrt{Q^2-\ell^2}}p^3-\frac{i}{2}  \sqrt{Q^2-\ell^2}\,p .
\end{align}

Comparing this result with the $-i \frac{c}{12} p^3 \delta_{p+n,0}$, we can read off the central charge
\begin{align}\label{RN:CL}
    c_L= \frac{ 6Q^4}{\sqrt{Q^2-\ell^2}}.
\end{align}

Let us continue to consider the temperature of Frolov-Thorne vacuum for black-bounce RN spacetime. Since the spacetime is spherically symmetric, the quantum field can be decomposed as 
\begin{align}
    \Phi =\sum_{\omega,l,\bar{m}} e^{-i \omega t} \mathcal{R}_{\omega l}(r) Y_{l \bar{m}}(\theta, \phi).
\end{align}
Here $\mathcal{R}_{\omega l}(r)$ is a function of radial coordinate $r$ with quantum number $\omega$ and $l$. $Y_{l \bar{m}}(\theta,\phi)$ is spherical harmonic function.

After tracing over the region inside the horizon, the density matrix of vacuum is 
\begin{align}
    \rho= e^{ \frac{- \hbar(\omega-q \Phi_{H})}{T_H}},
\end{align}
where $\Phi_H$ is electromagnetic potential difference between the horizon and infinity
\begin{align}
    \Phi_H=\chi^\mu A_\mu \bigg |_{r=\infty}- \chi ^\mu A_\mu \bigg|_{r=r+}= \frac{Q}{\sqrt{r_+^2+\ell^2}}.
\end{align}
Here $\chi =\partial_t$ is killing vector of horizon in spherically symmetric spacetime.

The Hawking temperature from \eqref{RN:TH} is 
\begin{equation}
    T_H=\frac{f'(r_+)}{4 \pi}= -\frac{r_+ \left(Q^2-m \sqrt{\ell^2+r_+^2}\right)}{2 \pi  \left(\ell^2+r_+^2\right)^2}.
\end{equation}

The left-moving and right-moving temperatures of Frolov-Thorne vacuum are
\begin{align}\label{RN:TLR}
    T_L=\frac{T_H}{1-\Phi_H}, \qquad T_R=\frac{T_H}{1+\Phi_H}. 
\end{align}
The details of deriving expressions $T_L$ and $T_R$ can be found in appendix \ref{Append}.

Taking the extremal limit $m \to Q$, the left-moving and right-moving temperatures turn to 
\begin{align}
    T_L&= \lim_{m \to Q} \frac{T_H}{1-\Phi_H}= \frac{ \sqrt{Q^2-\ell^2}}{2 \pi Q^2}, \label{RN:TL} \\     
    T_R&= \lim_{m \to Q} \frac{T_H}{1+\Phi_H}= 0.  
\end{align}

\subsection{Black hole entropy}

According to the holographic duality, the quantum states on near horizon region are identified with those of the left-moving part of the CFT. The central charge and temperature of CFT dual to the extremal black-bounce RN black holes are given by \eqref{RN:CL} and \eqref{RN:TL} 

Using the Cardy formula \cite{Cardy:1986ie}, the microscopic entropy of CFT can be obtained 
\begin{align}
    S_{\rm mic}&=\frac{\pi^2}{3}c_L T_L= \frac{\pi^2}{3}\times \frac{ 6Q^4}{\sqrt{Q^2-\ell^2}} \times  \frac{\sqrt{Q^2-\ell^2}}{2 \pi Q^2} \nonumber \\
    &= \pi Q^2 =S_{\rm  EBBRN}.
\end{align}
This result is consistent with the Bekenstein-Hawking entropy \eqref{RN:ES} of the extremal black-bounce RN black hole.

\section{Conclusion}\label{sec:conclu}

In this paper, we investigated the central charge and black hole entropy for regular extremal black-bounce spacetimes. Specifically, we analyzed the black-bounce counterparts of the Kerr, Kerr-Newman, and Reissner--Nordstr\"om spacetimes. In these geometries, the presence of the regularization parameter $\ell$ eliminates the singularity at $r=0$. Indeed, the curvature invariants such as $R$, $R^{\mu\nu}R_{\mu \nu}$, $R^{\mu \nu \alpha \beta}R_{\mu \nu \alpha \beta}$, and $C^{\mu \nu \alpha \beta}C_{\mu \nu \alpha \beta}$ are all finite and exhibit no divergence.

For the extremal black-bounce spacetimes considered here, the corresponding black hole entropy $S_{\text{BH}}$ can be readily obtained from the one-quarter area law, $S_{\rm BH}=\frac{1}{4}A_H$. This approach is standard in black hole thermodynamics, where the horizon area $A_H$ is computed via a codimension-two surface integral. Alternatively, the Kerr/CFT correspondence demonstrates that for the near horizon geometry of an extremal Kerr black hole, one can derive the central charge from the charge algebra of the asymptotic symmetry group and determine the left- and right-moving temperatures of the Frolov-Thorne vacuum. Subsequently, applying the Cardy formula $S=\frac{\pi^2}{3}c_L T_L$ of two-dimensional CFT yields the correct black hole entropy from a microscopic statistical perspective.

In this work, we adopted the Kerr/CFT approach to study the near horizon geometries of regular extremal black-bounce spacetimes. For the black-bounce Kerr and Kerr-Newman cases, the near horizon geometry exhibits an enhanced symmetry of SL$(2,\mathbb{R})\times \mathrm{U}(1)$. In contrast, for the black-bounce Reissner--Nordstr\"om case, the enhanced symmetry becomes SL$(2,\mathbb{R})\times \mathrm{SO}(3)$. We then analyzed the asymptotic symmetry groups and the corresponding charge algebras. For the black-bounce Kerr and Reissner--Nordstr\"om cases, the charge algebra consists solely of diffeomorphisms. It is worth emphasizing that in the black-bounce Reissner--Nordstr\"om case, we uplifted the 4D near horizon geometry to a 5D geometry by incorporating the $\mathrm{U}(1)$ gauge bundle of the electromagnetic field. Consequently, the diffeomorphisms in this case pertain to the 5D geometry rather than the 4D geometry. For the black-bounce Kerr-Newman case, the charge algebra includes both diffeomorphisms and $\mathrm{U}(1)$ gauge transformations. From these charge algebras, we extracted the central charges. In addition, the left- and right-moving temperatures were obtained by expanding the Boltzmann factor of the density matrix in the near horizon region. The final results for the central charge $c_L$, left-moving temperature $T_L$, microscopic entropy $S_{\rm mic}$, and black hole entropy $S_{\rm BH}$ for these extremal black-bounce spacetimes are summarized in Table \ref{tab:bh}.

\begin{table}[htbp]
\centering
\caption{Central charge $c_L$, left-moving Frolov-Thorne temperature $T_L$, microscopic entropy $S_{\rm mic}=\frac{\pi^2}{3}c_L T_L$, and black hole entropy $S_{\rm BH}=\frac{A_H}{4}$ for extremal black-bounce Kerr (BB-Kerr), black-bounce Kerr-Newman (BB-KN), and black-bounce Reissner-Nordstr\"om (BB-RN)  black holes.} \label{tab:bh}
\vspace{5mm}
\begin{tabular}{l c c c}
\toprule
    & \textbf{BB-Kerr} & \textbf{BB-KN} & \textbf{BB-RN} \\
\midrule
$\boldsymbol{c_L}$ 
& $\tfrac{12m^3}{\sqrt{m^2-\ell^2}}$ 
& $\tfrac{12m^2 \sqrt{m^2-Q^2}}{\sqrt{m^2-\ell^2}}$ 
& $\tfrac{6Q^4}{\sqrt{Q^2-\ell^2}}$ \\[8pt]
$ \boldsymbol{T_L}$ 
& $\tfrac{\sqrt{m^2-\ell^2}}{2m \pi}$ 
& $\tfrac{\sqrt{m^2-\ell^2}(2m^2-Q^2)}{4 \pi m^2 \sqrt{m^2-Q^2}}$ 
& $\tfrac{\sqrt{Q^2-\ell^2}}{2\pi Q^2 }$ \\[8pt]
$ \boldsymbol{S_{\rm mic}}$  
& $2m^2 \pi$ 
& $ (2m^2-Q^2)\pi $ 
& $ Q^2 \pi$ \\[6pt]
$ \boldsymbol{S_{\rm BH}}$ 
& $2m^2 \pi$ 
& $ (2m^2-Q^2)\pi $ 
& $ Q^2\pi  $ \\
\bottomrule
\end{tabular}
\end{table}

As is evident from the last two rows of Table \ref{tab:bh}, the microscopic entropy $S_{\rm mic}$ obtained from the Cardy formula agrees with the black hole entropy $S_{\rm BH}$ for these extremal black-bounce spacetimes. These results indicate that the Kerr/CFT approach remains valid for black hole spacetimes that are free of curvature singularities at $r=0$. This provides a viable framework for understanding black hole entropy from a microscopic statistical perspective.

\acknowledgments


\appendix

\section{$T_L$ and $T_R$ for Frolov-Thorne vacuum in black-bounce RN case}\label{Append}

In this part, we provide the details for deriving the formula \eqref{RN:TLR}, which is used in section \ref{sec:RNct}.

\subsection{Grand canonical ensemble}

Because the black-bounce RN black hole can exchange charge $Q$ with a thermal bath, the appropriate ensemble is the grand canonical ensemble with density matrix
\begin{equation}\label{App:DM}
    \rho = \exp\left[ -\beta_H \left(H-\Phi_H Q\right) \right],
\end{equation}
where $\beta_H=1/T_H$ is inverse Bekenstein-Hawking temperature and $Q$ is electric charge. The $\Phi_H$ is the electric potential evaluated at the horizon $r_+$.

Consider a charged scalar mode carrying energy $\omega$
and electric charge $q$,
\begin{equation}
\Psi \sim e^{-i\omega t}.
\end{equation}

Acting the density matrix \eqref{App:DM} on this mode is equivalent to perform the replacement $ H \rightarrow \omega$ and $ Q \rightarrow q$, then we obtain the thermal weight factor
\begin{equation}\label{eq:factor}
    \boxed{ e^{-\beta_H(\omega-q\Phi_H)} }.
\end{equation}
This is the RN analogue of the Kerr Frolov--Thorne factor $
e^{-\beta_H(\omega-m\Omega_H)}$

\subsection{Five-dimensional uplift}

Recall that to analyze the central charge of the extremal black-bounce RN metric, we combine the 4D near horizon geometry and $\mathrm{U}(1)$ electromagnetic field into a 5D geometry. The uplifted metric has the form
\begin{equation}
   \dd s_5^2 = \dd s_4^2 + \left(\dd y+A\right)^2,
\end{equation}
where $ y \sim y + 2\pi $ is a compact Kaluza--Klein circle.

In the five-dimensional case, a neutral quantum field may be expanded as
\begin{equation}
    \Psi \sim e^{-i\omega t + i k y}.
\end{equation}
Here $k$ is the conjugate momentum with respect to coordinate $y$.

Since $y$ is periodic, momentum $k$ should take the discrete integers $ k=n, n \in \mathbb{Z}$. After dimensional reduction, momentum along the Kaluza-Klein circle becomes electric charge in four dimensions, that is
\begin{equation}
    q=k. 
\end{equation}

Substituting this relation into the Frolov--Thorne factor \eqref{eq:factor} gives
\begin{equation}\label{eq:fac2}
    e^{-\beta_H\left( \omega-\Phi_H k \right)}.
\end{equation}

For the two-dimensional CFT, the thermal density matrix usually takes the form
\begin{equation}
    \rho_{\rm CFT} = \exp \left( -\frac{\omega_L}{T_L} -\frac{\omega_R}{T_R} \right).
\end{equation}

To identify the temperatures, we must rewrite the Frolov-Thorne factor $ \omega-\Phi_Hk$ in terms of left- and right-moving energies $\omega_L$ and $\omega_R$.

One can define
\begin{equation}
    \omega_L := \frac{1}{2}\left(\omega+k\right), \qquad \omega_R := \frac{1}{2} \left(\omega-k\right).
\end{equation}
These relations imply
\begin{equation}\label{eq:wk}
    \omega = \omega_L+\omega_R,\quad \text{and,}\quad k = \omega_L-\omega_R.
\end{equation}

\subsection{Left-moving and right-moving temperatures}

Substituting eq. \eqref{eq:wk} into the Frolov-Thorne exponent \eqref{eq:fac2} yields
\begin{align}
    \omega-\Phi_Hk &= (\omega_L+\omega_R) - \Phi_H(\omega_L-\omega_R) \nonumber \\
    &= (1-\Phi_H)\omega_L + (1+\Phi_H)\omega_R.
\end{align}

Therefore
\begin{equation}
    e^{-\beta_H(\omega-q\Phi_H)} = \exp \left\{ -\beta_H \left[ (1-\Phi_H)\omega_L + (1+\Phi_H)\omega_R \right] \right\}.
\end{equation}

Comparing this expression with
\begin{equation}
\exp \left( -\frac{\omega_L}{T_L} -\frac{\omega_R}{T_R} \right),
\end{equation}

one immediately obtains
\begin{equation}
    \frac1{T_L} = \beta_H(1-\Phi_H) \quad \text{and} \quad \frac1{T_R} = \beta_H(1+\Phi_H).
\end{equation}

Consequently, the left- and right-moving temperatures are obtained as
\begin{equation}
\boxed{ T_L = \frac{T_H}{1-\Phi_H}, \quad \text{and} \quad T_R = \frac{T_H}{1+\Phi_H} }.
\end{equation}


\begin{thebibliography}{99}

\bibitem{Bekenstein:1973ur}
J.~D.~Bekenstein,
{\em Black holes and entropy,}
Phys. Rev. D \textbf{7}, 2333 (1973).

\bibitem{Hawking:1975vcx}
S.~W.~Hawking,
{\em Particle Creation by Black Holes,}
Commun. Math. Phys. \textbf{43}, 199 (1975),
[erratum: Commun. Math. Phys. \textbf{46}, 206 (1976)].

\bibitem{Bardeen:1973gs}
J.~M.~Bardeen, B.~Carter, and S.~W.~Hawking,
{\em The Four laws of black hole mechanics,}  
Commun. Math. Phys. \textbf{31}, 161 (1973).

\bibitem{Hawking:1982dh}
S.~W.~Hawking and D.~N.~Page,
{\em Thermodynamics of Black Holes in anti-De Sitter Space,}
Commun. Math. Phys. \textbf{87}, 577 (1983).

\bibitem{Page:1993wv}
D.~N.~Page,
{\em Information in black hole radiation,}
Phys. Rev. Lett. \textbf{71}, 3743 (1993),
[arXiv:hep-th/9306083 [hep-th]].

\bibitem{Eisert:2008ur}
J.~Eisert, M.~Cramer, and M.~B.~Plenio,
{\em Area laws for the entanglement entropy - a review,}
Rev. Mod. Phys. \textbf{82}, 277 (2010),
[arXiv:0808.3773 [quant-ph]].

\bibitem{Harlow:2014yka}
D.~Harlow,
{\em Jerusalem Lectures on Black Holes and Quantum Information,}
Rev. Mod. Phys. \textbf{88}, 015002 (2016),
[arXiv:1409.1231 [hep-th]].

\bibitem{tHooft:1993dmi}
G.~'t Hooft,
{\em Dimensional reduction in quantum gravity,}
Conf. Proc. C \textbf{930308}, 284 (1993),
[arXiv:gr-qc/9310026 [gr-qc]].

\bibitem{Susskind:1994vu}
L.~Susskind,
{\em The World as a hologram,}
J. Math. Phys. \textbf{36}, 6377 (1995),
[arXiv:hep-th/9409089 [hep-th]].

\bibitem{Bigatti:1999dp}
D.~Bigatti and L.~Susskind,
{\em TASI lectures on the holographic principle,}
[arXiv:hep-th/0002044 [hep-th]].

\bibitem{Bousso:2002ju}
R.~Bousso,
{\em The Holographic principle,}
Rev. Mod. Phys. \textbf{74}, 825 (2002),
[arXiv:hep-th/0203101 [hep-th]].

\bibitem{Maldacena:1997re}
J.~M.~Maldacena,
{\em The Large $N$ limit of superconformal field theories and supergravity,}
Adv. Theor. Math. Phys. \textbf{2}, 231 (1998),
[arXiv:hep-th/9711200 [hep-th]].

\bibitem{Witten:1998qj}
E.~Witten,
{\em Anti de Sitter space and holography,}
Adv. Theor. Math. Phys. \textbf{2}, 253 (1998),
[arXiv:hep-th/9802150 [hep-th]].

\bibitem{Gubser:1998bc}
S.~S.~Gubser, I.~R.~Klebanov, and A.~M.~Polyakov,
{\em Gauge theory correlators from noncritical string theory,}
Phys. Lett. B \textbf{428}, 105 (1998),
[arXiv:hep-th/9802109 [hep-th]].

\bibitem{Aharony:1999ti}
O.~Aharony, S.~S.~Gubser, J.~M.~Maldacena, H.~Ooguri, and Y.~Oz,
{\em Large N field theories, string theory and gravity,}
Phys. Rept. \textbf{323}, 183 (2000),
[arXiv:hep-th/9905111 [hep-th]].

\bibitem{Strominger:1996sh}
A.~Strominger and C.~Vafa,
{\em Microscopic origin of the Bekenstein-Hawking entropy,}
Phys. Lett. B \textbf{379}, 99 (1996),
[arXiv:hep-th/9601029 [hep-th]].

\bibitem{Brown:1986nw}
J.~D.~Brown and M.~Henneaux,
{\em Central Charges in the Canonical Realization of Asymptotic Symmetries: An Example from Three-Dimensional Gravity,}
Commun. Math. Phys. \textbf{104}, 2076 (1986).


\bibitem{Strominger:1997eq}
A.~Strominger,
{\em Black hole entropy from near horizon microstates,}
JHEP \textbf{02}, 009 (1998),
[arXiv:hep-th/9712251 [hep-th]].

\bibitem{Sen:2007qy}
A.~Sen,
{\em Black Hole Entropy Function, Attractors and Precision Counting of Microstates,}
Gen. Rel. Grav. \textbf{40}, 2249 (2008),
[arXiv:0708.1270 [hep-th]].

\bibitem{Guica:2008mu}
M.~Guica, T.~Hartman, W.~Song, and A.~Strominger,
{\em The Kerr/CFT Correspondence,}
Phys. Rev. D \textbf{80}, 124008 (2009),
[arXiv:0809.4266 [hep-th]].


\bibitem{Cardy:1986ie}
J.~L.~Cardy,
{\em Operator Content of Two-Dimensional Conformally Invariant Theories,}
Nucl. Phys. B \textbf{270}, 186 (1986).

\bibitem{Anninos:2008fx}
D.~Anninos, W.~Li, M.~Padi, W.~Song, and A.~Strominger,
{\em Warped $AdS_3$ Black Holes,}
JHEP \textbf{03}, 130 (2009),
[arXiv:0807.3040 [hep-th]].

\bibitem{Frolov:1989jh}
V.~P.~Frolov and K.~S.~Thorne,
{\em Renormalized Stress-Energy Tensor Near the Horizon of a Slowly Evolving, Rotating Black Hole,}
Phys. Rev. D \textbf{39}, 2125 (1989).

\bibitem{Hartman:2008pb}
T.~Hartman, K.~Murata, T.~Nishioka, and A.~Strominger,
{\em CFT Duals for Extreme Black Holes,}
JHEP \textbf{04}, 019 (2009),
[arXiv:0811.4393 [hep-th]].

\bibitem{Lu:2008jk}
H.~Lu, J.~Mei, and C.~N.~Pope,
{\em Kerr/CFT Correspondence in Diverse Dimensions,}
JHEP \textbf{04}, 054 (2009),
[arXiv:0811.2225 [hep-th]].

\bibitem{Compere:2009dp}
G.~Compere, K.~Murata, and T.~Nishioka,
{\em Central Charges in Extreme Black Hole/CFT Correspondence,}
JHEP \textbf{05}, 077 (2009),
[arXiv:0902.1001 [hep-th]].

\bibitem{Bredberg:2009pv}
I.~Bredberg, T.~Hartman, W.~Song, and A.~Strominger,
{\em Black Hole Superradiance From Kerr/CFT,}
JHEP \textbf{04}, 019 (2010),
[arXiv:0907.3477 [hep-th]].

\bibitem{Castro:2010fd}
A.~Castro, A.~Maloney, and A.~Strominger,
{\em Hidden Conformal Symmetry of the Kerr Black Hole,}
Phys. Rev. D \textbf{82}, 024008 (2010),
[arXiv:1004.0996 [hep-th]].

\bibitem{Compere:2012jk}
G.~Comp{\`e}re,
{\em The Kerr/CFT Correspondence and its Extensions,}
Living Rev. Rel. \textbf{15}, 11 (2012),
[arXiv:1203.3561 [hep-th]].

\bibitem{Wang:2010qv}
Y.~Q.~Wang and Y.~X.~Liu,
{\em Hidden Conformal Symmetry of the Kerr-Newman Black Hole,}
JHEP \textbf{08}, 087 (2010),
[arXiv:1004.4661 [hep-th]].

\bibitem{Compere:2015bca}
G.~Comp{\`e}re, K.~Hajian, A.~Seraj, and M.~M.~Sheikh-Jabbari,
\emph{Wiggling Throat of Extremal Black Holes,}
JHEP \textbf{10}, 093 (2015),
[arXiv:1506.07181 [hep-th]].

\bibitem{Hajian:2017mrf}
K.~Hajian, M.~M.~Sheikh-Jabbari, and H.~Yavartanoo,
\emph{Extreme Kerr black hole microstates with horizon fluff,}
Phys. Rev. D \textbf{98}, 026025 (2018),
[arXiv:1708.06378 [hep-th]].

\bibitem{Stepanenko:2022gwy}
D.~Stepanenko and I.~Volovich,
\emph{Schwarzschild black holes, Islands and Virasoro algebra,}
Eur. Phys. J. Plus \textbf{138}, 688 (2023),
[arXiv:2211.03153 [hep-th]].

\bibitem{Volovik:2025yzr}
G.~E.~Volovik,
\emph{Thermodynamics of Kerr black hole: Tsallis{\textendash}Cirto composition law and entropy quantization,}
Pisma Zh. Eksp. Teor. Fiz. \textbf{123}, 515 (2026),
[arXiv:2509.00748 [gr-qc]].


\bibitem{Simpson:2018tsi}
A.~Simpson and M.~Visser,
{\em Black-bounce to traversable wormhole,}
JCAP \textbf{02}, 042 (2019),
[arXiv:1812.07114 [gr-qc]].

\bibitem{Mazza:2021rgq}
J.~Mazza, E.~Franzin, and S.~Liberati,
{\em A novel family of rotating black hole mimickers,}
JCAP \textbf{04}, 082 (2021),
[arXiv:2102.01105 [gr-qc]].

\bibitem{Franzin:2021vnj}
E.~Franzin, S.~Liberati, J.~Mazza, A.~Simpson, and M.~Visser,
{\em Charged black-bounce spacetimes,}
JCAP \textbf{07}, 036 (2021),
[arXiv:2104.11376 [gr-qc]].

\bibitem{Bardeen:1999px}
J.~M.~Bardeen and G.~T.~Horowitz,
{\em The Extreme Kerr throat geometry: A Vacuum analog of $AdS_2 \times S^2$,}
Phys. Rev. D \textbf{60}, 104030 (1999),
[arXiv:hep-th/9905099 [hep-th]].

\bibitem{Kunduri:2007vf}
H.~K.~Kunduri, J.~Lucietti, and H.~S.~Reall,
{\em Near-horizon symmetries of extremal black holes,}
Class. Quant. Grav. \textbf{24}, 4169 (2007),
[arXiv:0705.4214 [hep-th]].

\bibitem{Kunduri:2013gce}
H.~K.~Kunduri and J.~Lucietti,
{\em Classification of near-horizon geometries of extremal black holes,}
Living Rev. Rel. \textbf{16}, 8 (2013),
[arXiv:1306.2517 [hep-th]].

\bibitem{Ciambelli:2022vot}
L.~Ciambelli,
{\em From Asymptotic Symmetries to the Corner Proposal,}
PoS \textbf{Modave2022}, 002 (2023),
[arXiv:2212.13644 [hep-th]].

\bibitem{Ruzziconi:2019pzd}
R.~Ruzziconi,
{\em Asymptotic Symmetries in the Gauge Fixing Approach and the BMS Group,}
PoS \textbf{Modave2019}, 003 (2020),
[arXiv:1910.08367 [hep-th]].

\bibitem{Speziale:2025lkm}
S.~Speziale,
{\em GGI lectures on boundary and asymptotic symmetries,}
[arXiv:2512.16810 [hep-th]].

\bibitem{Barnich:2001jy}
G.~Barnich and F.~Brandt,
{\em Covariant theory of asymptotic symmetries, conservation laws and central charges,}
Nucl. Phys. B \textbf{633}, 3 (2002),
[arXiv:hep-th/0111246 [hep-th]].

\bibitem{Barnich:2007bf}
G.~Barnich and G.~Compere,
{\em Surface charge algebra in gauge theories and thermodynamic integrability,}
J. Math. Phys. \textbf{49}, 042901 (2008),
[arXiv:0708.2378 [gr-qc]].

\bibitem{Barnich:2003xg}
G.~Barnich,
{\em Boundary charges in gauge theories: Using Stokes theorem in the bulk,}
Class. Quant. Grav. \textbf{20}, 3685 (2003),
[arXiv:hep-th/0301039 [hep-th]].

\bibitem{Compere:2018aar}
G.~Comp{\`e}re and A.~Fiorucci,
{\em Advanced Lectures on General Relativity,}
[arXiv:1801.07064 [hep-th]].

\bibitem{Garousi:2009zx}
M.~R.~Garousi and A.~Ghodsi,
{\em The RN/CFT Correspondence,}
Phys. Lett. B \textbf{687}, 79 (2010),
[arXiv:0902.4387 [hep-th]].

\bibitem{Chen:2009ht}
C.~M.~Chen, J.~R.~Sun, and S.~J.~Zou,
{\em The RN/CFT Correspondence Revisited,}
JHEP \textbf{01}, 057 (2010),
[arXiv:0910.2076 [hep-th]].

\bibitem{Chen:2013rb}
B.~Chen, Z.~Xue, and J.~J.~Zhang,
{\em Note on Thermodynamic Method of Black Hole/CFT Correspondence,}
JHEP \textbf{03}, 102 (2013),
[arXiv:1301.0429 [hep-th]].

\end{thebibliography}
\end{document}